\documentclass[aps,pra,showkeys,superscriptaddress,11pt,notitlepage]{revtex4-1}
\usepackage{graphicx}  
\usepackage{amsmath,amsfonts,bbm,MnSymbol}   
\usepackage{ulem} 
\usepackage[breaklinks=true,colorlinks=true,linkcolor=blue,urlcolor=blue,citecolor=blue]{hyperref}
\usepackage{comment}
\usepackage{color}
\usepackage[makeroom]{cancel}
\pdfoutput=1
\usepackage{footnote}

\renewcommand{\[}{\begin{equation}}
\renewcommand{\]}{\end{equation}}

\begin{document}

\title{Dynamical Casimir effect in curved spacetime}

\author{Maximilian P. E. Lock}
\email{maximilian.lock12@imperial.ac.uk}
\address{Department of Physics, Imperial College, SW7 2AZ London, United Kingdom}
\address{Faculty of Physics, University of Vienna, Boltzmanngasse 5, 1090 Vienna, Austria}
\author{Ivette Fuentes}\thanks{Previously known as Fuentes-Guridi and Fuentes-Schuller.}
\affiliation{Faculty of Physics, University of Vienna, Boltzmanngasse 5, 1090 Vienna, Austria}
\affiliation{School of Mathematical Sciences, University of Nottingham, University Park, Nottingham NG7 2RD, United Kingdom}


\begin{abstract}
A boundary undergoing relativistic motion can create particles from quantum vacuum fluctuations in a phenomenon known as the dynamical Casimir effect. We examine the creation of particles, and more generally the transformation of quantum field states, due to boundary motion in curved spacetime. We provide a novel method enabling the calculation of the effect for a wide range of trajectories and spacetimes. We apply this to the experimental scenario used to detect the dynamical Casimir effect, now adopting the Schwarzschild metric, and find novel resonances in particle creation as a result of the spacetime curvature. Finally, we discuss a potential enhancement of the effect for the phonon field of a Bose-Einstein condensate.
\end{abstract}

\maketitle


\section{Introduction} \label{sIntro}
The dynamical Casimir effect (DCE) is the name given to the generation of particles due to changes in the mode structure of a quantum field, resulting either from the motion of a boundary constraining the field~\cite{moore1970quantum} or changing material properties of a medium containing the field~\cite{dodonov1993quantum} (see~\cite{dodonov2010current,dalvit2011fluctuations} for reviews). Physical implementations include photons generated by accelerated mirrors~\cite{moore1970quantum} (specifically due to changes in acceleration~\cite{fulling1976radiation,ford1982quantum,barton1993quantum}),  phononic excitations induced by changes in the external potential holding a Bose-Einstein condensate (BEC)~\cite{jaskula2012acoustic}, and photons generated by modulating the inductance of a superconducting quantum interference device (SQUID)~\cite{johansson2009dynamical}. The latter implementation has been demonstrated experimentally~\cite{wilson2011observation,lahteenmaki2013dynamical}.

The ability of relativistic motion (of an uncharged, apolar object) to produce nonclassical radiation from the vacuum~\cite{friis2012motion} provides a strong theoretical motivation to study the DCE. In addition, the quantum nature of the radiation produced has led to investigations of its utility for a wide variety of quantum information tasks such as entanglement generation~\cite{friis2013entanglement,johansson2013nonclassical,busch2014quantum,aron2014steady,felicetti2014dynamical,rossatto2016entangling}, the generation of quantum discord~\cite{sabin2015quantum}, the production of cluster states for quantum computation~\cite{bruschi2016towards}, the performance of multipartite quantum gates~\cite{friis2012quantum,bruschi2013relativistic}, quantum steering~\cite{sabin2015generation} and quantum communication~\cite{benenti2014dynamical}.

The DCE lies at the interface between quantum mechanics and relativity, and can be treated within the framework of quantum field theory in curved spacetime (QFTCS)~\cite{fulling1976radiation,birrell1984quantum}. This allows relativistic considerations such as the analogy between the DCE and the radiation emitted by a collapsing star~\cite{davies1977radiation}, or the effect of motion on a quantum clock in the famous twin paradox scenario~\cite{lindkvist2014twin}. In the context of quantum cosmology, one can consider particle creation due to some expanding boundaries of the universe as a kind of DCE~\cite{brevik2000dynamical}, and likewise for graviton creation in string theoretic models with moving branes~\cite{durrer2007dynamical}. The DCE due to the motion of a single boundary near a black hole has been studied from a thermodynamic perspective~\cite{unruh1982acceleration}, calculating the energy flux in 1+1 dimensions and showing it to be negative. Subsequent work has discussed the prospect that this negative energy flux is unphysical~\cite{walker1985negative} or unobservable~\cite{ottewill1988radiation}. More recently, boundary motion in a static curved spacetime was investigated in~\cite{celeri2009action}, considering a cavity with a single mirror moving briefly over a short distance. However, a general description of the transformation of quantum field states due to the DCE in curved spacetime has remained an open problem. Here, we provide a novel method which enables the latter to be calculated, and is applicable to a wide range of scenarios. This allows the consideration of new experiments manifesting both quantum and general-relativistic features, and testing QFTCS. We describe the effect of a finite period of motion through curved spacetime on a quantum field state contained between two boundaries (e.g.\ an optical cavity). We first briefly review previous approaches to the calculation of the DCE (section~\ref{sPrevApproaches}) before describing some relevant aspects of QFTCS (section~\ref{sFramework}), which we then use as a framework to obtain the main result (section~\ref{sMainResult}). We reconsider the experimental scenario used to detect the DCE in section~\ref{sExample}, where we find novel particle-creation resonances when one includes spacetime curvature, as well as agreement with previous results. Then, in section~\ref{sBECexample}, we briefly sketch the possibility of an enhancement of the effect in a BEC system.


\section{Previous approaches to calculating the DCE} \label{sPrevApproaches}
Methods for calculating the DCE can be broadly separated into two categories (with some overlap). Hamiltonian methods such as those described in~\cite{razavy1985quantum,dodonov2010current,haro2007physically} allow, for example, the consideration of the finite refractive index of the mirrors~\cite{barton1993quantum,barton1996peculiarities}, and the resistive forces acting on them due to the created particles~\cite{dalvit2011fluctuations,jaekel1992motional}. On the other hand, one can consider the solution of the field equations subject to some externally imposed boundary trajectories, an approach in which we can employ QFTCS, and which is therefore more suited to relativistic considerations. We adopt the latter approach here. The first calculations of the DCE were carried out in this manner, exploiting conformal transformations (which leave the field equation unchanged from its inertial-coordinate form) to some coordinates in which the boundaries are stationary~\cite{moore1970quantum,candelas1977vacuum,fulling1976radiation}. This reveals the difficulty of maintaining a particle interpretation during the boundary motion (see the introductory discussion in~\cite{fulling1976radiation}), a problem which is equally present for quantum fields in arbitrary spacetimes~\cite{birrell1984quantum}. A number of exact solutions using this approach in flat spacetime, in the case of a single moving mirror are given in~\cite{good2013time}.

A distinct variant of this conformal approach is developed in~\cite{bruschi2013mode}, which takes a view ``local'' to an observer at the center of a cavity, which undergoes some time-dependent proper acceleration. The boundary trajectories are set such that they are at a constant distance in the instantaneous Rindler frame corresponding to the observer's proper acceleration at a given moment in time. One can then calculate the effect of a finite period of acceleration on the quantum field inside the cavity, not with a single conformal coordinate transformation, but rather by integrating through a continuum of them.

For arbitrary boundary motion, it is not possible to find a conformal transformation between inertial coordinates and some coordinates in which the boundaries are stationary. Given trajectories for which such a transformation cannot be found, one can instead seek the solution to the field equation in terms of ``instantaneous''	 mode solutions~\cite{razavy1985quantum,ji1997production}, and then use some approximations particular to the given trajectories to solve the resulting infinite set of coupled differential equations. This can then be used to connect solutions before a finite period of motion to solutions afterwards - effectively a scattering problem. In appendix~\ref{a1}, we demonstrate the use of this method for a particular subset of trajectories, where the problem can be solved without using their explicit form, and we show that this coincides with the results obtained using the novel method outlined in section~\ref{sMainResult}. For general boundary trajectories however, the former method cannot be applied without further approximations, unlike the method we present here, where the only condition is that the boundary motion is slow with respect to the speed of light.


\section{Framework} \label{sFramework}
We consider a cavity containing a massless real scalar field $\Phi$, which can be used to approximate the electromagnetic field when polarization effects are negligible~\cite{friis2013scalar}, or to describe phononic excitations in certain BEC setups~\cite{fagnocchi2010relativistic}. We present our results in $1+1$ dimensions, though the same arguments can be applied equally well with an arbitrary number of spatial dimensions as long as the Klein-Gordon equation is separable in space and time, as is the case for static spacetimes (see appendix~\ref{a2}).

First let us consider flat spacetime, with inertial coordinates $(t,x)$. The Klein-Gordon equation subject to the stationary boundary conditions ${\Phi(t,x=x_{1})=}{\Phi(t,x=x_{2})=0}$, admits the solutions
\begin{equation} \label{eModeSolns}
\phi_{m}(t,x) = N_{m} e^{-i \omega_{m} t} \sin \left[ \omega_{m} (x-x_{1}) \right]
\end{equation}
and their complex conjugates, where $N_{m}=1 / \sqrt{m \pi}$ is a normalization constant, $\omega_{m} = m \pi /  L$ are the mode frequencies (with $m =1,2,3,\ldots$ labeling the mode), $L =  x_{2}-x_{1}$ is the cavity length, and we have chosen ${c=1}$. The inner product between solutions is given by~\cite{birrell1984quantum}
\begin{equation} \label{eInnerProd}
(\varphi,\chi) = -i \int_{x_{1}}^{x_{2}} \mathrm{d}x \left[ \varphi \left( \partial_{t} \chi^{*} \right) - \chi^{*} \left( \partial_{t} \varphi \right) \right].
\end{equation}
The mode solutions are orthonormal in the sense that $(\phi_{m},\phi_{n})=\delta_{mn}, \,$ $(\phi_{m}^{*},\phi_{n}^{*})=-\delta_{mn}$ and $(\phi_{m},\phi_{n}^{*})=0$. They are then associated with particles via bosonic annihilation and creation operators $a_{m}$ and $a_{m}^\dag$, and the vacuum state and Fock space are defined in the usual way~\cite{birrell1984quantum}, with the total field operator given by
\begin{equation} \label{eTotalField}
\Phi(t,x) = \sum_{m} \left[ a_{m} \phi_{m}(t,x) +  a_{m}^\dag \phi_{m}^{*}(t,x)  \right] .
\end{equation}
A linear transformation from one set of mode solutions to another is known as a Bogoliubov transformation. Such transformations arise, for example, when considering changes in coordinate system (such as Lorentz boosts, or transformations between inertial and accelerated observers), or as a result of spacetime dynamics~\cite{birrell1984quantum}, or when describing the action of a unitary process whose generating Hamiltonian is an at-most-second-order polynomial in the creation and annihilation operators (examples include the displacement, squeezing and beam-splitting operations, and any other Gaussian operation in quantum optics)~\cite{weedbrook2012gaussian}. In section~\ref{sMainResult}, we will use a Bogoliubov transformation to describe the effect of a finite period of boundary motion. Gathering the mode solutions in equation~(\ref{eModeSolns}) into a column vector $\Psi = [\phi_{1},\phi_{2},\ldots,\phi_{1}^{*},\phi_{2}^{*},\ldots]^{T}$, we can write the Bogoliubov transformation to some new set of solutions $\tilde{\Psi}$ as a matrix equation $\tilde{\Psi}=S \Psi$, with (in block matrix form)
\begin{equation} \label{eSmatrices}
S=
\begin{bmatrix}
\alpha & \beta \\
\beta^{*} & \alpha^{*}
\end{bmatrix},\ \quad SKS^\dag=K,\ \quad K:=
\begin{bmatrix}
I & 0 \\
0 & -I
\end{bmatrix},
\end{equation}
where $\alpha_{mn}=(\tilde{\phi}_{m},\phi_{n})$ and $\beta_{mn}=-(\tilde{\phi}_{m},\phi_{n}^{*})$ are known as the Bogoliubov coefficients,
and the middle equation (containing the so-called Bogoliubov identities) ensures the orthonormality of the transformed solutions. The composition of multiple transformations is calculated by multiplying the corresponding matrices. Equations~(\ref{eSmatrices}) identify $S$ as an element of a complex representation of a real symplectic group~\cite{Arvind1995a}. Given the Bogoliubov coefficients, one can compute the corresponding transformation of the creation and annihilation operators, and therefore the transformation of a given quantum state. In particular, the $\beta_{mn}$ quantify particle creation due to the transformation. For example, starting with a vacuum state, the average number of particles in mode $m$ after a Bogoliubov transformation is given by $\mathcal{N}_{m}= \sum_{n} \left\vert \beta_{nm} \right\vert^{2}$.

We now consider the cavity to be embedded in some curved spacetime, and assume that the latter admits a timelike Killing vector field in the region of interest, so that we can construct a well-defined Hilbert space from the solutions to the field equation~\cite{wald1994quantum}. It is always possible to find some coordinate system in which the metric is conformally flat~\cite{birrell1984quantum}, and consequently the Klein-Gordon equation takes the same form as in inertial coordinates in flat space. Letting $(t,x)$ now denote the conformally flat coordinates, one finds that the framework above, in particular equations~(\ref{eModeSolns}) to (\ref{eTotalField}), holds.


\section{Main result} \label{sMainResult}
We now derive the Bogoliubov transformation corresponding to a finite period of cavity motion in curved spacetime, before and after which, the field solutions have the same form as in equation~\ref{eModeSolns}. To do this, we make the assumption that, for boundary motion much slower than the speed of light, an infinitesimal time-step can be described as the combination of a displacement effect and a pure phase-evolution of stationary mode solutions. In other words, we assume that the boundaries are moving slowly enough that they are effectively stationary on the timescale of a massless particles' reflection. Then, in the same vein as~\cite{bruschi2013mode}, a differential equation for the total transformation can be derived. We first define a matrix of frequencies $\Omega:=\text{diag}\left( \omega_{1}, \omega_{2}, \ldots, -\omega_{1}, -\omega_{2}, \ldots \right)$. To make explicit the dependence on the boundary conditions, let us write this as $\Omega(x_{1},x_{2})$ and the stationary mode solutions in equation~(\ref{eModeSolns}) as $\phi_{m}(t,x;x_{1},x_{2})$. Using the same phase convention as~\cite{bruschi2012voyage}, we can write the transformation from the solutions at $\phi_{m}(t,x;x_{1},x_{2})$ to solutions at $\phi_{m}{(t+\delta t,x;x_{1}+\delta x_{1},x_{2}+ \delta x_{2})}$ as $\exp \left[ i \Omega(x_{1}+\delta x_{1},x_{2}+\delta x_{2} ) \delta t \right] S_{\delta}$, where $S_{\delta}$ is composed of the Bogoliubov coefficients obtained by taking inner products between $\phi_{m}(t,x;x_{1},x_{2})$ and $\phi_{m}(t,x;{x_{1}+\delta x_{1}},{x_{2}+ \delta x_{2}})$. If we now consider motion for some finite time $t$, and denote the corresponding transformation matrix by $S(t)$, then composing transformations gives $S(t+\delta t){=\exp \left[ i \Omega(x_{1}+\delta x_{1},x_{2}+\delta x_{2} ) \delta t \right] S_{\delta}(\delta x_{1},\delta x_{2} ) S(t)}$, leading to the differential equation
\begin{equation} \label{eFullDiffEqu}
\frac{d S}{dt} = \left[  i \Omega + M^{(1)} \frac{d x_{1}}{dt} + M^{(2)} \frac{d x_{2}}{dt}  \right] S
\end{equation}
where 
\begin{equation} \label{eGenerators}
M^{(j)}=
\begin{bmatrix}
A^{(j)} & B^{(j)} \\
B^{(j)*} & A^{(j)*}
\end{bmatrix}, \; \;
A^{(j)}_{mn}:=\left( \frac{\partial \phi_{m}}{\partial x_{j}} , \phi_{n} \right) \, \text{ and } B^{(j)}_{mn}:=-\left( \frac{\partial \phi_{m}}{\partial x_{j}} , \phi_{n}^{*} \right),
\end{equation}
and with $j=1,2$. The formal solution to equation~(\ref{eFullDiffEqu}) can be written using the time-ordered exponential. Consider motion between $t=0$ and $t=T$, and following~\cite{celeri2009action} let us define $\Theta(t)=\int_{0}^{t} \mathrm{d} t' \Omega(t')$. Now, since $\Omega$ is a diagonal matrix we can solve equation~(\ref{eFullDiffEqu}) in a manner analogous to the interaction picture of quantum mechanics, giving
\begin{equation} \label{eTexpSoln}
S(T)= e^{i \Theta(T)} \mathcal{T} \mathrm{exp} \left[ \int_{0}^{T} \mathrm{d} t  \sum_{j=1}^{2} e^{- i \Theta(t)}  M^{(j)}  e^{ i \Theta(t)} \frac{d x_{j}}{dt} \right] .
\end{equation}

Given our assumption that the coordinate velocities of the boundaries $\frac{d x_{j}}{dt}$ are small (with respect to the speed of light) throughout the motion, we can use the Dyson series to express the time-ordered exponential in equation~(\ref{eTexpSoln}) to first order in $\frac{d x_{j}}{dt}$. The Bogoliubov coefficients are then
\begin{subequations} \label{eBogos}
\begin{align}
\alpha_{mn} &= e^{i \int_{0}^{T} \mathrm{d} t \omega_{m}(t)} \left\lbrace \delta_{mn} + \sum_{j=1}^{2}  \int_{0}^{T} \mathrm{d}t \, A_{mn}^{(j)} e^{-i \int_{0}^{t} \mathrm{d}t' \left[ \omega_{m}(t') - \omega_{n}(t') \right]} \, \frac{d x_{j}}{dt}   \right\rbrace   		\\
\label{eBogosBeta} \beta_{mn} &= e^{i \int_{0}^{T} \mathrm{d} t \omega_{m}(t)} \sum_{j=1}^{2} \int_{0}^{T} \mathrm{d}t \, B_{mn}^{(j)} e^{-i \int_{0}^{t} \mathrm{d}t' \left[ \omega_{m}(t') + \omega_{n}(t') \right]} \, \frac{d x_{j}}{dt}.
\end{align}
\end{subequations}
These coefficients satisfy the identities given by equations~(\ref{eSmatrices}). In appendix~\ref{a1} we compare this result with that obtained using a different method (described in section~\ref{sPrevApproaches}), to consider the subset of trajectories where $L$ is constant, in which case one can approximately solve the field equations to find the Bogoliubov coefficients without using the specific functional form of the trajectory. We show that the answer thus obtained coincides with equations~(\ref{eBogos}). For trajectories where $L$ is not constant, one cannot always apply the method described in section~\ref{sPrevApproaches}. A generalisation of equations~\ref{eBogos} to arbitrarily many spatial dimensions is given in appendix~\ref{a2}. In the next section we apply our method introduced here to one such trajectory, and recover known results in the flat-space limit.

One can see the physical role played by each term in equations~(\ref{eBogos}); in $\alpha_{mn}$ they give respectively the total phase accrued and mode-mixing due to the motion of each boundary, while the two terms of $\beta_{mn}$ correspond to particle creation by the two moving boundaries. Integrating by parts, one can see a correspondence between equations~(\ref{eBogos}) above and equations~(6) in~\cite{bruschi2013mode}. 

We exploit the simplicity of the Klein-Gordon equation in conformally flat coordinates, but trajectories in coordinates more natural to a given problem can be mapped to ones in the conformally flat coordinates. Furthermore, the temporal coordinate $t$ is used as a bookkeeping coordinate, which can be related to the proper time of an observer in the usual way. Both of these points are illustrated in the following example.


\section{Example: an oscillating boundary in the presence of a massive body} \label{sExample}
We now consider a scenario where, in the presence of a stationary, spherically symmetric, massive body, one boundary is fixed while the other oscillates in the direction radial to the body. This can be used to model an experiment on the surface of the Earth, ignoring the Earth's rotation. The DCE due to boundary oscillation in flat spacetime is a well-studied problem, e.g.~\cite{dodonov1995photon,lambrecht1996motion,ji1997production}, and indeed the oscillating-boundary scenario was used to observe the DCE experimentally~\cite{wilson2011observation}. There, one finds a resonance in the creation of particles when the boundary oscillates at the sum-frequency of two modes. This resonance has been examined theoretically in a weak gravitational field using a short-time approximation~\cite{celeri2009action}. Here, we use our novel method to find simple expressions for the $\beta_{mn}$ coefficients, revealing further resonances due to the spacetime curvature.

We describe the spacetime curvature due to the massive body using the Schwarzschild metric, given by $\mathrm{d}s^{2}=-f(r) \mathrm{d}t^{2}+\frac{1}{f(r)}\mathrm{d}r^{2}$ with $f(r):=1-r_{S}/r$, where $r_{S}=2 G M$ is the Schwarzschild radius of the body. One can relate the proper time $\tau$ of, for example, some stationary experimenter at a radial distance $r_{e}$ to the bookkeeping time coordinate $t$ using $\tau=\sqrt{f(r_{e})} t$. We consider one boundary at $r_{1}=r_0$ to be fixed and the other boundary at $r_2(t)=\left(r_{0}+L_{0}\right)\left[1+\delta(t)\right]$ to move from $t=0$ to $t=T$ such that there is a sinusoidal oscillation of the proper length, i.e.
\begin{equation} \label{ePropLengthOscil}
L_{p}(t) = L_{p,0} + \tilde{A} \sin \left( \nu t \right).
\end{equation}
with $L_{p,0} = \int_{r_0}^{r_{0}+L_{0}} \frac{\mathrm{d}r}{f(r)}$. For simplicity of presentation, it is assumed the boundary returns to its initial position at $t=T$, i.e. $\nu T = p \pi$ for some $p \in \mathbb{N}$. Assuming the oscillation amplitude to be much smaller than the distance to the centre of the gravitating body, one finds
\begin{equation}
r_2(t)=r_{0}+L_{0} + A \sin (\nu t)    \qquad  \text{ with }   \qquad  A=\frac{\sqrt{f\left( r_{0}+L_{0} \right)}}{\sqrt{f\left( r_{0}+L_{0} \right)}+\frac{r_{s}}{2\left( r_{0}+L_{0} \right)}} \tilde{A},
\end{equation} 
to first order in $\delta(t)$.  We further assume $\varepsilon := A/L_{0}<<1$ and $\varepsilon>>r_{S}/r$, which are easily satisfied in experiments at the Earth's surface. We will work to first order in $r_{S}/r$ and second order in $\varepsilon$. For reference, in SQUID-based DCE experiments one can achieve a fractional change of the (effective) length as large as $\sim 0.1$~\cite{johansson2010dynamical,wilson2011observation}, and at the surface of the Earth, we have $r_{S}/r\sim10^{-9}$.

To find the $\beta_{mn}$ coefficients, we use equation~\ref{eBogosBeta}, employing the so-called tortoise coordinate $x:=r+r_{S} \ln \left\vert \frac{r}{r_{S}}-1 \right\vert$, wherein the metric is conformally flat, giving
\begin{equation} \label{eOscillatingBeta}
\beta_{mn}=  e^{i \omega_{n}T} \varepsilon \nu  \, \sqrt{ \omega_{m} \omega_{n} } \frac{f(r_{0})}{f(r_{0}+L_{0})} \left\lbrace i \frac{ (-1)^{p} - e^{i \left( \omega_{m} + \omega_{n} \right) T} }{\left( \omega_{m} + \omega_{n} \right)^{2} - \nu^{2}} + \frac{A \, r_{S} }{(r_{0}+L_{0})^{2} } \, \frac{\nu}{ \omega_{m} + \omega_{n}} \,  \frac{e^{i \left( \omega_{m} + \omega_{n} \right) T}-1}{  \left( \frac{ \omega_{m} + \omega_{n} }{2}\right)^{2}-\nu^{2} }  \right\rbrace .
\end{equation}
The first term in equation~(\ref{eOscillatingBeta}) persists in the limit of zero curvature (i.e. $r_{S} \to 0$), and exhibits the familiar resonance for a driving frequency of $\nu= {\omega_{m}} {+ {\omega_{n}}}$. The second term gives a novel contribution due to curvature, with its own resonance at the subharmonic ${\nu= \frac{1}{2} \left({ \omega_{m}} + {\omega_{n}} \right)}$~\footnote{To avoid a potential confusion, we emphasize that this half-wavelength resonance is distinct from the fact that, when driving the mirror at $\nu=\omega_{m}+\omega_{n}$, one obtains a peak in the \textit{output spectrum} of an initially-empty cavity at $\nu/2$ (or the nearest frequencies to that, if $m+n$ is odd).}, though this is strongly suppressed by the factor of $A r_{S} / (r_{0}+L_{0})^2$. Including higher orders of $\varepsilon$ and $r_{S}/r$ into the calculation of $\beta_{mn}$, one finds further resonances. These resonances are a result of the nonlinear relationship between the proper length and the length in the conformally flat coordinate $x_{2}-x_{1}$. Stated in more physical terms, the length relevant to the experimenter, the proper length, differs nontrivially from the appropriate notion of length for a massless particle (sometimes called the ``radar length''~\cite{rindler2006relativity}), and this difference depends on the curvature of the spacetime, quantified in this case by $r_{S}$. A single-frequency sinusoidal modulation of the proper length corresponds to a complex motion in the conformally-flat length. This complex motion can be written as a weighted sum of sinusoidal terms, each with a different frequency, and each leading to a new resonance. Conversely, one could imagine the experimenter contriving a complex modulation of the proper length such that the motion in the $x$-coordinate is exactly sinusoidal, in which case only the standard resonance would remain.

Considering a driving frequency at a resonance of the first term, i.e. $\nu = {\omega_{q}} + {\omega_{r}}$ for some $q$ and $r$, and then considering the regime $\nu T>>1$, one obtains ${\left\vert \beta_{mn} \right\vert^{2}} {= \frac{1}{4} \left(1- 2 \frac{L_{0} r_{s}}{r_{0}^{2}} \right) mn \left( \varepsilon \frac{f(r_{0}) \pi}{L_{0}} T \right)^{2} \delta_{m+n,q+r}}$. We thus find a curvature-induced reduction in particle number, as noted in~\cite{celeri2009action}, and we recover equation~(4.5) of~\cite{ji1997production} in the flat-spacetime limit. This reduction is in line with the physical interpretation of the novel resonances noted above; the sinusoidal driving from the perspective of the laboratory in not sinusoidal in the physically relevant coordinate $x$.

If we now drive the boundary at the novel resonance, $\nu= \frac{1}{2} \left( {\omega_{q}} + {\omega_{r}} \right)$ for some $q$ and $r$, the coefficient for which the curvature-dependent contribution is largest is
\begin{equation} \label{eOscBetaSubharm}
\beta_{qr} = i  e^{i \omega_{q}T} \, \varepsilon\frac{\sqrt{qr} f(r_{0})^{2} }{f(r_{0}+L_{0})}  \left[ -\frac{2}{3} \frac{1-(-1)^{p}}{f(r_{0}) \left( q+r \right) } + \varepsilon \frac{\pi}{8} \frac{r_{S} T}{ f(r_{0}+L_{0}) (r_{0}+L_{0})^{2}} \right].
\end{equation}
From equation~(\ref{eOscBetaSubharm}), one can see that it is in principle possible to conduct an experiment for long enough that the curvature-dependent contribution dominates, since in the $\beta_{mn}$ for modes other than $q$ and $r$ (equation~(\ref{eOscillatingBeta})), the value of $T$ only serves to set the phases. Taking the massive body to be the Earth, and considering the parameters used in the SQUID setup of~\cite{lindkvist2014twin} at the surface, one finds that the observation of this resonance would take $10^{17}$ times longer than observation of the usual parametric resonance, which is evidently impracticable. Instead of a photonic DCE, we can consider phononic excitations of a BEC. There are cases where relativistic effects too small to detect with an optical cavity may be brought into an observable regime with a BEC setup~\cite{bruschi2014phonon,bruschi2014testing}, and by preparing a suitable probe state and measuring its transformation due to the motion, one can profit from the increased sensitivity afforded by quantum metrology~\cite{ahmadi2014quantum,ahmadi2014relativistic}. We are currently studying the feasibility of such a scheme, and we present the first steps of this study in section~\ref{sBECexample}. Nonetheless, the in-principle detectability of the curvature contribution is an incentive for further study of this and other trajectories.


\section{Amplification of the effect in a BEC} \label{sBECexample}

As noted in section~\ref{sIntro}, modulating the trapping potential of a BEC affects the mode structure of the phonons, leading to a DCE. This has been implemented in~\cite{jaskula2012acoustic}, for example, where correlated phonon pairs are produced by sinusoidal modulation of the trap potential. We now give a cursory argument for the possibility of using this platform to detect the effect described in section~\ref{sExample}. In the following, we reintroduce the speed of light $c$ in order to compare magnitudes.

Using a relativistic mean-field description of the BEC, it has been shown that under certain conditions (see equations~62 of~\cite{fagnocchi2010relativistic}, and appendix~A of~\cite{bruschi2014testing}), the phonon field obeys the Klein-Gordon equation with the effective metric:
\begin{equation} \label{eEffMet}
\mathfrak{g}_{\mu\nu} = \rho \frac{c}{c_s} \left[ g_{\mu\nu} + \left( 1- \frac{c_{s}^2}{c^2} \right)  \frac{v_{\mu}v_{\nu}}{c^2} \right]
\end{equation}
where $\rho$, $c_{s}$ and $v^{\mu}$ are respectively the density, speed of sound and four-velocity of the mean-field (see~\cite{fagnocchi2010relativistic} for detailed definitions). We assume the confining potential to be a ``box trap''~\cite{gaunt2013bose}, so that the phonon field effectively resides in a cavity, and we consider a 1+1D Schwarzschild spacetime as in the previous section. We now seek the coordinates which are conformally flat with respect to the effective metric $\mathfrak{g}_{\mu\nu}$. As a first, rather coarse, approximation we assume the BEC mean-field to be completely static, so that $c_{s}$ and $\rho$ are constant and $v^{\mu}=(c/ \sqrt{f(r)},0)$. The latter results from the condition $v_{\mu}v^{\mu}=-c^{2}$. Applying these approximations to equation~\ref{eEffMet}, one finds that the conformally flat coordinates $(t,x)$ are given by is the usual Schwarzschild time coordinate and $x:=\frac{c}{c_{s}} \left[ r+r_{s} \ln \left( \frac{r}{rs} - 1 \right) \right]$. The latter is simply a rescaling of the tortoise coordinate used in section~\ref{sExample} by a factor of $\frac{c}{c_{s}}$. Taking the speed of sound to be on the order of $1\text{cm s}^{-1}$~\cite{jaskula2012acoustic}, we have $\frac{c}{c_{s}}\sim 10^{10}$. Considering the form of equation~\ref{eBogosBeta}, we see that the $\beta_{mn}$ are also scaled by this factor, greatly enhancing the magnitude of the effect, and perhaps enabling the detection of the novel resonances predicted in section~\ref{sExample}.


\section{Discussion}
We have given a novel, simple method which allows the calculation of the DCE due to boundary motion in curved spacetime. We have presented the method in $1+1$ dimensions for simplicity, with an extension to higher dimensions given in appendix~\ref{a2}. As well as giving some general formulas for the Bogoliubov transformation of the field state, we have considered the experimental scenario used to observe the DCE~\cite{wilson2011observation}, found a novel resonance in particle creation if one includes spacetime curvature, and briefly considered the possibility of amplifying the effect in a BEC. Our method can therefore be used to consider a range of experiments manifesting quantum and general relativistic effects. It can also be applied to extend investigations into the effect of motion on quantum properties such as entanglement, for example~\cite{bruschi2012voyage}, to analyze the effect of spacetime curvature. We now note some limitations of our approach, and some possible extensions.

By prescribing the boundary trajectories, we ignore the backreaction (and therefore resistive force) on the mirrors due to particle creation. In light of the additional curvature-dependent terms in the example above, we see that such a backreation will be affected by the presence of gravity, increasing or decreasing the ``quantum vacuum friction''~\cite{dalvit2011fluctuations,davies2005quantum} resisting the motion of an object through spacetime.

Our assumption of perfectly-reflecting boundaries implies that the purity of the field state inside the cavity is unaffected by the motion. Relaxing this assumption would allow a consideration of coupling between intra-cavity modes and global ones, and the resulting loss of purity. The result would by a fully-relativistic description of decoherence induced by non-inertial motion, including the effect of gravity.

It would be of interest to see if our approach can be modified to consider asymptotically-static motion of a single boundary through curved space. This would require the approach described in sections~\ref{sFramework} and~\ref{sMainResult} to be adapted for the continuous-spectrum case. One could then compare trajectories in curved spacetime with known flat-spacetime results such as those described in~\cite{good2013time} in order to investigate how the presence of spacetime curvature affects those results.

Finally, we note that our analysis of the BEC implementation was a sketch for the purpose of showing the possible benefit of using such a platform, and that the subject deserves a much fuller treatment, with serious attention to experimental details. It seems likely that the static-BEC assumption will need to be refined, and one may need to consider the effect of both the static spacetime curvature and the dynamic trapping potential on the spatiotemporal dependence of the mean-field properties (see~\cite{robertson2017controlling} for an example of the latter effect).

\section*{Acknowledgments}
The authors would like to thank Dominik \v{S}afr\'{a}nek, Tupac Bravo, David Jennings, Luis C. Barbado and David E Bruschi for useful discussions and comments. M. P. E. L. acknowledges support from the EPSRC via the Controlled Quantum Dynamics CDT (EP/G037043/1).


\appendix

\section{Comparison with Bogoliubov coefficients obtained using an ``instantaneous basis'' of mode solutions: the constant-length case}\label{a1}
Here, we consider the subset of cavity trajectories for which $x_{2}(t)-x_{1}(t)=L$, a constant, and demonstrate a different way of calculating the Bogoliubov coefficients (used for example in~\cite{ji1997production}), showing that this coincides with the result given in the main article. 

As in our main article, the coordinate system in which the metric is conformally flat is denoted $(t,x)$, and the positions of the cavity walls in these coordinates are given by $x_{1}(t)$ and $x_{2}(t)$.  In the same manner as~\cite{yuce2008dynamical}, we move to new coordinates $(t,q)$ with $q(t,x):=x-x_{1}(t)$, and thus
\begin{equation}
\partial_{t}^{2} - \partial_{x}^{2} \to \partial_{t}^{2} + \left( \dot{q}^2 - 1 \right) \partial_{q}^{2} + \ddot{q} \partial_{q} + 2 \dot{q} \partial_{q} \partial_{t}
\end{equation}
where a dot denotes the derivative with respect to $t$. To incorporate the assumption of low-velocity, i.e. $|\dot{q}|<<1$, we assume that we can write $\dot{q}(t)= \eta \dot{y}(t)$ for some $y(t)$ and $\eta<<1$. To first order in $\eta$, the Klein-Gordon equation is then
\begin{equation}\label{eKG}
\left[ \partial_{t}^{2} - \partial_{q}^{2} + \ddot{q} \partial_{q} + 2 \dot{q} \partial_{q} \partial_{t} \right] \Phi(t,q) = 0
\end{equation}
with the (now time-independent) boundary conditions $\Phi(t,0)=\Phi(t,L)=0$. We seek solutions $\varphi_{m}(t,q)$ to Equation~\ref{eKG} using an ``instantaneous basis'' consisting of the spatial part of the stationary-cavity solutions given in Equation~\ref{eModeSolns}:
\begin{equation} \label{eSolns}
\varphi_{m}(t,q) = \sum_{p} Q_{mp}(t) N_{p} \sin \left( \omega_{p} q    \right)
\end{equation}
For a cavity which is stationary for $t \leq 0$ and $t \geq T$, we have the conditions
\begin{subequations}
\begin{align}
\varphi_{m}(t \leq 0,q) &= N_{m} e^{-i \omega_{m} t} \sin \left( \omega_{m} q    \right)  \\
\varphi_{m}(t \geq T,q) &= \sum_{p} \left\lbrace N_{p} \left( \tilde{\alpha}_{mp} e^{-i \omega_{p} t} + \tilde{\beta}_{mp} e^{i \omega_{p} t} \right) \sin \left( \omega_{p} q    \right) \right\rbrace ,
\end{align}
\end{subequations}
i.e.
\begin{subequations} \label{eTimeConds}
\begin{align}
Q_{mn}(t \leq 0) &= e^{-i \omega_{m} t} \delta_{mn}  \label{eTimeConds0}   \\
Q_{mn}(t \geq T) &= \tilde{\alpha}_{mn} e^{-i \omega_{n} t} + \tilde{\beta}_{mn} e^{i \omega_{n} t}, \label{eTimeCondsT}
\end{align}
\end{subequations}
where $\tilde{\alpha}_{mn}$ and $\tilde{\beta}_{mn}$ are the coefficients encoding how the post-motion solutions can be written in terms of the pre-motion ones. We now insert the solutions in Equation~\ref{eSolns} into Equation~\ref{eKG} and integrate out the spatial part to obtain an infinite set of coupled differential equations for the $Q_{mn}(t)$. To do this, we use the following identities
\begin{subequations}
\begin{align}
\int_{0}^{L} \mathrm{d}q \, N_{m} \sin \left( \omega_{m} q    \right) N_{n} \sin \left( \omega_{n} q    \right) &= \frac{\delta_{mn}}{2 \omega_{n}} \\
\int_{0}^{L} \mathrm{d}q \, N_{m} \cos \left( \omega_{m} q    \right) N_{n} \sin \left( \omega_{n} q    \right) &= g_{mn}
\end{align}
\end{subequations}
where
\begin{equation} \label{egmn}
g_{mn} := \begin{cases} 
   0 & \text{for } m=n \\
   \frac{\sqrt{mn} \left[ 1-(-1)^{m+n} \right] }{(m+n)(m-n) \pi}       & \text{for } m \neq n.
  \end{cases}
\end{equation}
We thus obtain
\begin{equation}
\ddot{Q}_{mn} + \omega_{n}^{2} Q_{mn} = 2 \omega_{n} \sum_{p} \left(  2 \dot{q} \, \dot{Q}_{mp} + \ddot{q} \, Q_{mp} \right) g_{pn}
\end{equation}
Seeking solutions up to first order in $\eta$ we write $Q_{mn} = Q_{mn}^{(0)} + \eta Q_{mn}^{(1)} $, and hence obtain equations for the zero and first-order parts:
\begin{subequations} \label{eSepOrdDiffEqun}
\begin{align}
\ddot{Q}_{mn}^{(0)} + \omega_{n}^{2} Q_{mn}^{(0)} &= 0   \\
\ddot{Q}_{mn}^{(1)} + \omega_{n}^{2} Q_{mn}^{(1)} &= 2 \omega_{n} \sum_{p} \left(  2 \dot{y} \, \dot{Q}_{mp}^{(0)} + \ddot{y} \, Q_{mp}^{(0)} \right) g_{pn}.
\end{align}
\end{subequations}
The condition given in Equations~\ref{eTimeConds0} then becomes
\begin{equation} \label{eTimeConds02}
Q_{mn}^{(0)}(t \leq 0) = e^{-i \omega_{m} t} \delta_{mn} \qquad \text{and} \qquad Q_{mn}^{(1)}(t \leq 0)=0.
\end{equation}
Assuming continuity of $Q_{mn}(t)$ and $\dot{Q}_{mn}(t)$ at $t=0$, we can use these conditions and their derivatives to obtain
\begin{subequations}
\begin{align}
Q_{mn}^{(0)}(t) &= e^{-i \omega_{m} t} \delta_{mn}  \\
Q_{mn}^{(1)}(t) &= g_{mn} \int_{0}^{t} \mathrm{d} s \left[   \left( \omega_{m}+\omega_{n} \right) e^{-i \left( \omega_{m} - \omega_{n} \right) s} e^{-i \omega_{n} T} - \left( \omega_{m}-\omega_{n} \right) e^{-i \left( \omega_{m} + \omega_{n} \right) s} e^{i \omega_{n} T}  \right] \dot{y}(s).
\end{align}
\end{subequations}
Defining $A_{mn} := - \left( \omega_{m}+\omega_{n} \right) g_{mn}$ and $B_{mn} := \left( \omega_{m}-\omega_{n} \right) g_{mn}$, we have
\begin{equation}
Q_{mn}^{(1)}(t) = - \int_{0}^{t} \mathrm{d} s \left[  A_{mn} e^{-i \left( \omega_{m} - \omega_{n} \right) s} e^{-i \omega_{n} T} + B_{mn} e^{-i \left( \omega_{m} + \omega_{n} \right) s} e^{i \omega_{n} T}  \right] \dot{y}(s).
\end{equation}
We then obtain the $\tilde{\alpha}_{mn}$ and $\tilde{\beta}_{mn}$ by using the condition given in Equations~\ref{eTimeCondsT} and continuity at $t=T$
\begin{subequations}
\begin{align}
\tilde{\alpha}_{mn} &= \delta_{mn} +A_{mn} \int_{0}^{T} \mathrm{d} t \,  e^{-i \left( \omega_{m} - \omega_{n} \right) t} \,  \dot{x}_{1}(t) \\
\tilde{\beta}_{mn} &= B_{mn} \int_{0}^{T} \mathrm{d} t \,  e^{-i \left( \omega_{m} + \omega_{n} \right) t} \, \dot{x}_{1}(t)
\end{align}
\end{subequations}
These correspond to the transformation from the pre-motion mode solutions to the post-motion solutions, where both sets of solutions are evaluated at $t=T$. To transform from pre-motion solutions evaluated at $t=0$ to post-motion solutions evaluated at $t=T$, we ``undo'' the phase evolution of the pre-motion solutions from $t=0$ to $t=T$, giving the Bogoliubov coefficients
\begin{subequations} \label{eInstModeSolnsBogos}
\begin{align} 
\alpha_{mn} &= e^{i \omega_{m} T} \left[ \delta_{mn} +A_{mn} \int_{0}^{T} \mathrm{d} t \,  e^{-i \left( \omega_{m} - \omega_{n} \right) t} \,  \dot{x}_{1}(t) \right] \\
\beta_{mn} &= e^{i \omega_{m} T} B_{mn} \int_{0}^{T} \mathrm{d} t \,  e^{-i \left( \omega_{m} + \omega_{n} \right) t} \, \dot{x}_{1}(t),
\end{align}
\end{subequations}
To show that Equations~\ref{eInstModeSolnsBogos} coincide with the method given in the main article, we first calculate the matrices $A^{(j)}$ and $B^{(j)}$ defined in Equations~\ref{eGenerators}:
\begin{subequations} \label{eAmnBmn}
\begin{alignat}{2}
\label{As} A^{(1)}_{mn} &= \begin{cases} 
   0 & \text{ for } m=n \\
   \frac{(-1)^{m+n} \sqrt{\omega_{m}\omega_{n}}}{L(\omega_{m}-\omega_{n})}       & \text{ for } m \neq n
  \end{cases} &\qquad \quad A^{(2)}_{mn} &= \begin{cases} 
   0 & \text{ for } m=n \\
   - \frac{\sqrt{\omega_{m}\omega_{n}}}{L(\omega_{m}-\omega_{n})}       & \text{ for } m \neq n
  \end{cases} \\	\nonumber & & \\
\label{Bs} B^{(1)}_{mn} &= - \frac{(-1)^{m+n} \sqrt{\omega_{m}\omega_{n}}}{L(\omega_{m} + \omega_{n})}  & B^{(2)}_{mn} &= \frac{ \sqrt{\omega_{m}\omega_{n}}}{L(\omega_{m} + \omega_{n})} , 
\end{alignat}
\end{subequations}
where we have used the time-independence of the inner product~\cite{birrell1984quantum}. We note the similarity between these and Equations~22 of~\cite{bruschi2014phonon}. Now, since $\dot{L}=0$, we have $\dot{x}_{1} = \dot{x}_{2}$, and then solving Equation~\ref{eFullDiffEqu} to first order in $\dot{x}_{1} $ one obtains
\begin{subequations} \label{eNewMethSolns}
\begin{align}
\alpha_{mn} &= e^{i \omega_{m} T} \left[ \delta_{mn} +\left( A_{mn}^{(1)}+A_{mn}^{(2)} \right) \int_{0}^{T} \mathrm{d} t \,  e^{-i \left( \omega_{m} - \omega_{n} \right) t} \,  \dot{x}_{1}(t) \right] \\
\beta_{mn} &= e^{i \omega_{m} T} \left( B_{mn}^{(1)}+B_{mn}^{(2)} \right) \int_{0}^{T} \mathrm{d} t \,  e^{-i \left( \omega_{m} + \omega_{n} \right) t} \, \dot{x}_{1}(t).
\end{align}
\end{subequations}
Comparing this to Equations~\ref{eInstModeSolnsBogos}, we can see that the Bogoliubov coefficients obtained using the two methods coincide if $A=A^{(1)}+A^{(2)}$ and $B=B^{(1)}+B^{(2)}$, which indeed holds, as we can see by considering Equation~\ref{egmn} and Equations~\ref{eAmnBmn}.

\section{Generalisation to \textit{D} spatial dimensions}\label{a2}
As noted in the main text, it is not necessary to restrict ourselves to $1+1$ dimenensions. Consider now the case of a static spacetime with $D$ spatial dimensions. There then exists some coordinates $(t, \overrightarrow{x})$ (where $\overrightarrow{x}$ has components $x^{k}$ and $k=1,2,\ldots,D$) such that the Klein-Gordon equation is separable, i.e. one can seek solutions in the form $\phi_{m}(t,\overrightarrow{x})=T_{m}(t)X_{m}(\overrightarrow{x})$ (where the label $m$ is no longer a single number). In this case, $\partial_{t}$ is a timelike Killing vector, and (ignoring normalisation) we have $T_{m}(t)=e^{-i \omega_{m} t}$ for some $\omega_{m}$. As in section~\ref{sMainResult}, we write the solutions as $\phi_{m}(t,\overrightarrow{x};\overrightarrow{x}_{1},\overrightarrow{x}_{2})$, where $\overrightarrow{x}_{1}$ and $\overrightarrow{x}_{2}$ denote the boundary positions in $D$ spatial dimensions, and we can now follow exactly the same procedure as before, with $S_{\delta}$ now comprised of inner products between $\phi_{m}\left( t,\overrightarrow{x};\overrightarrow{x}_{1},\overrightarrow{x}_{2} \right)$ and $\phi_{m}\left( t,\overrightarrow{x};\overrightarrow{x}_{1}+\delta \overrightarrow{x}_{1},\overrightarrow{x}_{2}+\delta \overrightarrow{x}_{2} \right)$, and the total transformation then satisfies the multi-dimensional generalisation of equation~\ref{eFullDiffEqu}
\begin{equation} \label{eFullDiffEquMultiDim}
\frac{d S}{dt} = \left[  i \Omega + M^{(1)}_{k} \frac{d x^{k}_{1}}{dt} + M^{(2)}_{k} \frac{d x^{k}_{2}}{dt}  \right] S
\end{equation}
where the sum over $k$ is implicit, and
\begin{equation} \label{eGeneratorsMultiDim}
M^{(j)}_{k}=
\begin{bmatrix}
A^{(j)}_{k} & B^{(j)}_{k} \\
B^{(j)*}_{k} & A^{(j)*}_{k}
\end{bmatrix}, \; \;
\left(A^{(j)}_{k}\right)_{mn}:=\left( \frac{\partial \phi_{m}}{\partial x^{k}_{j}} , \phi_{n} \right) \, \text{ and } \left(B^{(j)}_{k}\right)_{mn}:=-\left( \frac{\partial \phi_{m}}{\partial x^{k}_{j}} , \phi_{n}^{*} \right),
\end{equation}
with $j=1,2$.


\bibliographystyle{naturemag}
\bibliography{DCEinCS_BiBTeX}

\end{document}